\documentstyle[11pt,epsfig]{article}
\title{\bf Self-Relative (or Machian) Information\\
$S_{BH}=\frac{1}{4}A_{BH}-\frac{3}{2}\log A_{BH}$}
\author{Nima Khosravi\thanks{email: nima@ipm.ir,
n-khosravi@sbu.ac.ir}\\
{\small {\it Department of Physics, Shahid Beheshti University, G.
C., Evin, Tehran 19839, Iran }}}

\begin{document}
\maketitle
\begin{abstract}
The entropy-area relation of black holes is one of the important
results of theoretical physics. It is one of the few relations that
is used to test theories of quantum gravity in the absence of any
experimental evidence. It states that $4 \times \ell_P^2$ is the
fundamental area that holds \textit{one} bit of information.
Consequently, a question arises: why $4 \times \ell_P^2$ and not $1
\times \ell_P^2$ is the fundamental holder of \textit{one} bit of
information? In any case it seems the latter choice is more natural.
We show that this question can be answered with a more explicit
counting of the independent states of a black hole. To do this we
introduce a method of counting which we name self-relative
information. It says that a bit alone does not have any information
unless it is considered near other bits. Utilizing this approach we
obtain the correct entropy-area relation for black holes with $1
\times \ell_P^2$ as the fundamental holder of \textit{one} bit of
information. This method also predicts, naturally, the existence of
logarithmic corrections to the entropy-area relation.
\vspace{5mm}\newline PACS: 04.70.Dy (Quantum aspects of black holes,
evaporation, thermodynamics), 04.60.-m (Quantum gravity)
\end{abstract}
\maketitle
\section{Introduction}
Black holes are very important in classical relativity as well as
quantum gravity. In classical physics the notions of black hole and
big bang play a crucial role in understanding their singular
behavior \cite{GRbooks}. It is generally believed that the
gravitational field becomes dominant near the singularities
resulting in breaking down of classical general relativity. And as a
consequence, the quantum effects of gravity become worthy of
consideration. As mentioned above, the behavior of singularities in
classical general relativity in one hand and the foundations of
quantum mechanics on the other, may lead to the resolution of the
question of singularities in quantum general relativity theory
\cite{QGbooks,STbooks}. From another viewpoint, in the theory of
everything, the final theory should resolve all the existing
problems in current theories such as singular behaviors. For
example, string theory as a candidate for the theory of
everything\footnote{Or the other approaches of quantum gravity as a
part of the everything theory.} should present some clear ideas on
black holes. In addition, lack of any direct experimental data in
quantum gravity regime causes an ambiguity on the correctness of
proposed theories. Until such experimental evidence, the theoretical
work plays the crucial role of verifying the correctness of such
theories. One of these theoretical evidences is the comparison of
different methods. The most trusted method is the semi-classical
analysis\footnote{Since, at least, this method coincides with the
classical results in the appropriate limits.}. Fortunately, for the
black hole behavior there is some semi-classical analysis and
predictions. The agreement of these predictions with the predictions
of the proposed theories is an important\footnote{Maybe the most
important.} sign for the correctness of those theories. In the
following, we will focus on the problem of black hole entropy.

The entropy of black holes is one of the very interesting problems
in the theoretical physics. In calculations related to black holes,
the fundamental constants ($c$, $\hbar$ and $G$) appear and tie to
each other and interestingly, this is exactly the realm of quantum
gravity. The more interesting feature is that the resulting entropy
can be deduced in the absence of any full quantum gravity
\cite{hawking bh}. As mentioned above, it is very essential to check
quantum gravity candidates. Because there exists a result in the
quantum gravity regime that can be a tester for theories of quantum
gravity e.g. string theory or canonical quantum gravity and so on.
The story of black hole entropy began with the possible
contradiction between the existence of black holes and the second
law of thermodynamics. Avoiding this contradiction results in an
analogy between thermodynamical quantities and black hole's
properties \cite{bardeen}. It is worth to mention that the similar
four laws of black hole mechanics are just some analogies in
classical regime when introduced in \cite{bardeen}. To understand
better the nature of these analogies and consequently find an
interpretation for them it is necessary to enter quantum phenomena.
The mentioned analogies make a generalization in the second law of
thermodynamics that a black hole has an entropy. This entropy of a
black hole is proportional to its area, $S_{BH}\propto A$, as
conjectured by Bekenstein \cite{bekenstein}. The factor of
proportionality is fixed by Hawking \cite{hawking bh} such that
$S_{BH}=\frac{1}{4}A$. As mentioned before, Hawking did the
calculations with a semi-classical method. Nowadays different
approaches show the same result for the entropy of black holes e.g.
in string theory \cite{string bh} and also in canonical quantum
gravity \cite{loop bh}. In addition, these quantum theories of the
gravity predict a logarithmic correction term with a method
dependent pre-factor.

On the other hand, the discrete structure of geometry is commonly
believed as a consequence of quantum gravity \cite{smolin}. This
kind of structure makes it possible to find the entropy of a black
hole due to counting the possibilities \cite{wheeler} and calculate
the entropy by Shanon law $S\propto \log P$
\cite{shanon}\footnote{\label{footnote}In \cite{shanon} the basis of
the logarithmic function is $2$ and it seems it is related to the
definition of probability and possibility. This may resolve a $\log
2$ factor in the final entropy-area relation.}, where $P$ is the
number of possible states. In more details, one can have an area
proportional to the minimum area\footnote{$\ell_P^2$ is the only
natural choice for the minimum area.}, $A=N \ell_P^2$ where $\ell_P$
is the Planck length. Letting two possible values for each
fundamental area\footnote{This means that the fundamental area can
only hold two different bits $0$ or $1$. The calculations do not
depend to this special proposition and the method works properly for
general cases, i.e. $d$-level systems, as it will be shown.} results
in $P=2^N$ possible states. By Shanon law the entropy becomes $S
\propto N$ and then $S\propto A$ or $S=k \frac{A}{\ell_P^2}$. As has
been mentioned in the literature there is no evidence in this
approach to find the proportionality constant $k$
\cite{strominger}\footnote{In his lecture, Strominger expresses some
difficulties in quantum gravity. Here it is worth to quote the first
hint in his lecture: ``If we tile the horizon with Planck-sized
cells, and assign one degree of freedom to each cell, then the
entropy, which is extensive, will go like the area. This suggests
that the microstates can be described as living on the horizon
itself. The hard part is to naturally get the $\frac{1}{4}$ from
such a picture." Deducing this $\frac{1}{4}$ is the main part of
current work.}. Comparison of this information based method and
other approaches \cite{hawking bh} determines that $k=\frac{1}{4}$.
It means that $4\times \ell_P^2$ holds \textit{one} bit of
information, a $0$ or a $1$. In our opinion this is a bizarre
result, since naturally $1 \times \ell_P^2$ should hold \textit{one}
bit of information. In this paper we will try to show that each $1
\times \ell_P^2$ holds \textit{one} bit of information while still
the same standard relation between area and entropy of a black hole
is valid. In the following we will introduce the notion of
self-relative information to establish physically meaningful
information. Then we will use the suggested method of counting to
obtain the black hole entropy-area relation. In addition we will
show that the procedure imposes a logarithmic correction term,
naturally. We will close the paper with concluding remarks.

\begin{figure}[th]
\centerline{\includegraphics[width=10cm]{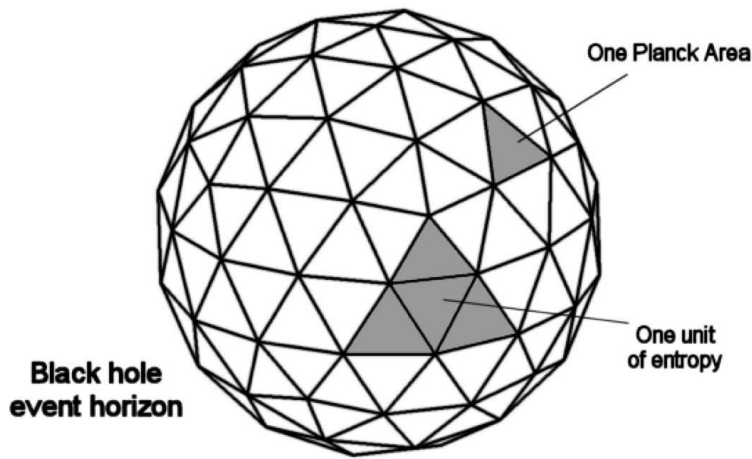}}
\caption{\label{fig1}\footnotesize This figure shows the standard
interpretation of entropy-area relation for a black hole,
schematically. In other words, the area is decomposed to fundamental
areas, $1 \times \ell_P^2$, but the unit of entropy (information) is
$4 \times \ell_P^2$. We have shown in the body of the paper that
both of area and entropy (information) units are same and equal to
$1 \times \ell_P^2$ in the context of self-relative information
procedure. The figure is borrowed from
http://www.scholarpedia.org/article/Image:BHentropyF1.jpeg.}
\end{figure}

\section{Self-Relative Information}
To commence this section it is worth mentioning briefly some points
about the information-based viewpoint on black hole properties. As
stated above, there is an analogy between black hole properties and
thermodynamical quantities. On the other hand the thermodynamics can
be seen by statistical mechanics' tools. In this form all the
macroscopic thermodynamical quantities have a microscopic
interpretation. For example the entropy shows variety of microstates
that are constrained by a given macroscopic condition. Consequently,
black hole macroscopic properties, i.e. its mass, angular momentum
and electric charge, can be illustrated by some microscopic states.
The next step is finding the microscopic states for a black hole
which is done in the context of string theory, loop quantum gravity
and also heuristic ways e.g. tiling the area of a black hole by the
fundamental areas\footnote{Historically, this way of thinking has
been heuristic but nowadays there is a physical interpretation e.g.
from loop quantum gravity.}. Attaching different states to each
fundamental area makes this viewpoint very similar to information
theory which contains a sequence of bits. This similarity results in
natural usage of information theory concepts in black hole theory.
One of these concepts is the definition of the entropy in
information theory which can be characterized by Shanon law as
mentioned previously. In this approach large entropy produces a
large amount of information \cite{horo}. To meet this concept two
dual approaches have been considered, a subjective picture versus an
objective one \cite{horo}. In the subjective picture, information is
known by the sender but is unknown for the receiver. However, in the
objective viewpoint, the information is known for the receiver. To
go further we will work in the subjective picture which is
characterized by Shanon's entropy. Now let us introduce a question
to enter more details.

How much information exist in a sequence of $0$'s and $1$'s? Or how
much information can be stored in $N$ bits of memory filled by $0$
and $1$? The straightforward answer is that since each bit has two
different values then a sequence of $N$ bits has $2^N$
distinguishable states. At first glance, it is correct but there is
some ambiguities. To be clearer, let us try to answer how does
computer understand what is the meaning of a sequence of bits? The
computer compares the given sequence with its database to say for
example in seven bit ASCII code, $1000101 1001110 1010100 1010010
1001111 1010000 1011001$ means ENTROPY. To make this correspondence,
access to the ASCII code table is necessary and without the table it
is impossible. It means that to find the meaning of a sequence of
bits, a dictionary is an essential requirement. As another example,
in cryptography when the data is sent to someone, he must have the
relevant database to understand the content of the message. Now what
about the cases for which we do not have any access to the
appropriate dictionary? What about the number of possible states on
the area of a black hole?

As mentioned above, inexistence of a proper dictionary makes
understanding of a given sequence of bits impossible. Since if we
cannot study the meaning of a sequence, it is not physically
understandable then we must ignore it. Now let we utilize the
counting program to deduce entropy of a black hole. What is in front
of us? Similar to the above discussions we do not have any
dictionary to translate the data on the area of a black hole in an
appropriate way. The essential question appears naturally, how one
can solve this problem not only for a black hole but generally? We
will show in a sequence of bits some information exist even in the
absence of a dictionary and call it self-relative information. The
heart of the idea is that when there is no definite translator in
nature then nature must choose a coding procedure which makes
self-access possible. In a sequence of bits without a dictionary
each bit does not contain any meaningful information but its
relative distance to other bits in the sequence can contain
understandable and obtainable information. This idea is totaly in
agreement with the belief that there is no preferred observer in the
universe and everything is relative\footnote{One can assume the
dictionary exists but it has been lost. In this case there is two
philosophically different choices, one stays on to find the
dictionary and the other utilizes the relative information. We pick
the second one in our discussion that is more usual in theoretical
physics specially after special and general relativity theory.}. In
other words similar to the idea of Mach for geometry\footnote{It
says that there is no geometry in the presence of vacuum.} and the
heart of general relativity \cite{QGbooks}\footnote{The general
relativity is a background independent theory i.e. only the relative
quantities are physical quantities.} there is no information for a
sequence of only $0$'s\footnote{Since there is no information in $0$
or $1$ and only their difference is a physical object, similar to
the sign of electric charges, then it is true for a sequence of only
$1$'s.}. It is easy to see that according to this kind of thinking
on the notion of information, the amount of information in a
sequence of bits is smaller in comparison to the standard viewpoint.
Again, we stress that this new definition for the physical
information is based on the relative relations of bits in the
sequence e.g. relative distances. Now let us to count the number of
physically understandable states on the area of a black hole in the
next section.

\section{Black Hole Entropy}
To calculate the entropy of a black hole we will use the counting
method. In this method primarily we suppose that the structure of
area is quantized and each quantum of area holds a bit of
information. As mentioned before this method with the above
assumptions breaks down because of the lack of the factor
$\frac{1}{4}$ in the entropy-area relation. We show that this factor
can be reproduced if one only attends to the understandable states.
In other words, one must count only those states which are
distinguishable. It is important to say that in this procedure we
must note that we have not allowed any ambiguities to surface. The
last phrase is essential in our calculations and we will see it in
more details in the following. The existence of any ambiguities
result in disability to recover information from a given sequence of
bits. So, if we believe in recovery of information of nature then
separating ambiguous sequences will be crucial and necessary not
only in calculation but also in philosophy.

To distinguish different sequences two approaches exist, the first
one is to compare two different sequences and state their
equivalence or independence, the equivalent second method is
constructing the independent sequences and then counting them. The
second approach is more straightforward and we chose it here. Let
the black hole be a sphere with area $A_{BH}=N^2 \ell_P^2$. In the
first step all the bits contains $0$ and we want to add $1$'s step
by step and count the independent possibilities. Since sketching a
sphere is not simple we use a circle but we know that the boundaries
are imaginary. To change the first bit from $0$ to $1$ how many
choices we have? The answer is $N^2$! But no, all our choices are
equivalent because we cannot distinguish them, so for the first $1$
we have only one choice which we name it point $A$ in figure 2. What
about the second one? $N^2-1$ choices? In this step different bits
cause different sequences because of the existence of the first $1$
(i.e. a point with a label, $A$). The relative distance between $A$
and the second choice makes different sequences distinguishable. And
since choosing the second point $B$ with a distance $d$ from $A$ is
not sensitive to the direction of equators pass through $A$ then all
these equators become equivalent and picking each of them up makes
no independent sequence, figure 3. To reach the independent
sequences one must pick up one of them and choose $B$ with a
distance $d$ from $A$, figure 3. There exists still an ambiguity
because of two choices for $B$ on a line passing through $A$. To
remove this ambiguity we can pick up both choices and make the third
choice bearing in mind that $B_1$ and $B_2$ are equivalent, figure
4. It is important that these different $B$'s are not
distinguishable. So picking up both of them eliminates the worry
about the ambiguity. Obviously choosing a point on an equator of a
sphere with circumference $A_{BH}=N^2 \ell_P^2$ has $\sim N$
possibilities. It is interesting to mention that the ignorance on
such ambiguities reduces the $\sim N^4$ choices for both of the
first points to $\sim N$. We will show that such dividing by $\sim
N^3$ results in logarithmic correction to the entropy-area relation
of a black hole. Turning to the third choice, it is the most
important choice to get the correct $\frac{1}{4}$ factor in
entropy-area relation. Suppose we want to choose the third point,
$C$, in relative distances $d_1$ and $d_2$ with respect to $A$ and
$B$. We will continue the discussions in the two following
subsections, in the first one we will show how $\frac{1}{4}$ appears
naturally in the self-relative information proposal and in the
second subsection, the appearance of the logarithmic correction
term.
\begin{figure}[th]
\centerline{\includegraphics[width=12cm]{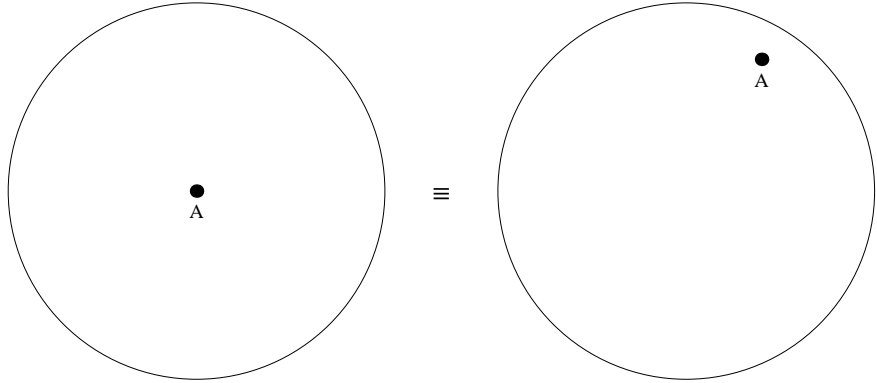}}
\caption{\label{fig1}\footnotesize To pick the first point, there is
no differences between the points on the area. We will continue with
the left figure without any loss of generality. It is worth
mentioning that the first point, $A$, cannot understand the
dimensionality of the area. This feature is crucial to reach to
logarithmic correction term.}
\end{figure}

\begin{figure}[th]
\centerline{\includegraphics[width=12cm]{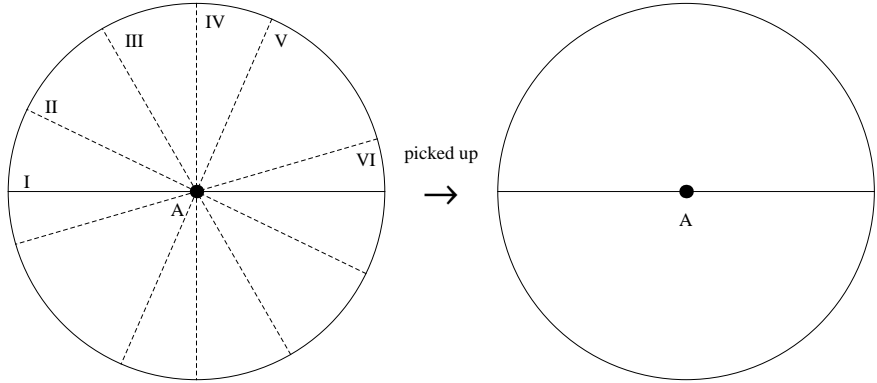}}
\caption{\label{fig1}\footnotesize The second point is in a definite
distance to the first point, $A$. This additional second point and
the first point can present only one dimension. But all the lines
through $A$ are same as each other due to the symmetry of the area.
We picked the horizontal one without any loss of generality. Note,
blindness of two points to the second dimension of the area plays a
crucial role in appearing the logarithmic correction term.}
\end{figure}

\begin{figure}[th]
\centerline{\includegraphics[width=6cm]{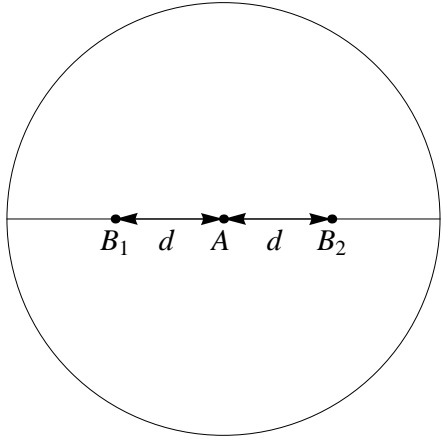}}
\caption{\label{fig1}\footnotesize To ignore any ambiguity we pick
up two equivalent points for the second choice with same distances,
$d$ with respect to $A$.}
\end{figure}

\begin{figure}[th]
\centerline{\includegraphics[width=6cm]{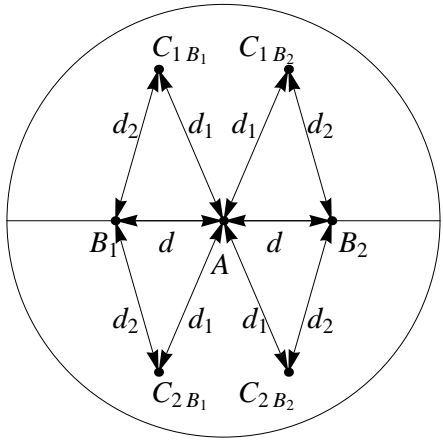}}
\caption{\label{fig1}\footnotesize The critical point is the third
point since three points can understand the dimensionality of the
area completely. There is four choices with $d_1$ and $d_2$
distances with respect to $A$ and $B$ respectively. Note that there
is two equivalent points, $B_1$ and $B_2$. After this step there is
four equivalent sets of points, $\{A,B_1,C_{1B_1}\}$,
$\{A,B_1,C_{2B_1}\}$, $\{A,B_2,C_{1B_2}\}$ and $\{A,B_2,C_{2B_2}\}$.
To pick up the fourth choice there is a unique point due to each
set. The same is true for the latter choices that results in four
absolutely same copies of a picture on the area see figure 5.}
\end{figure}

\begin{figure}
\begin{tabular}{ccc} \epsfig{figure=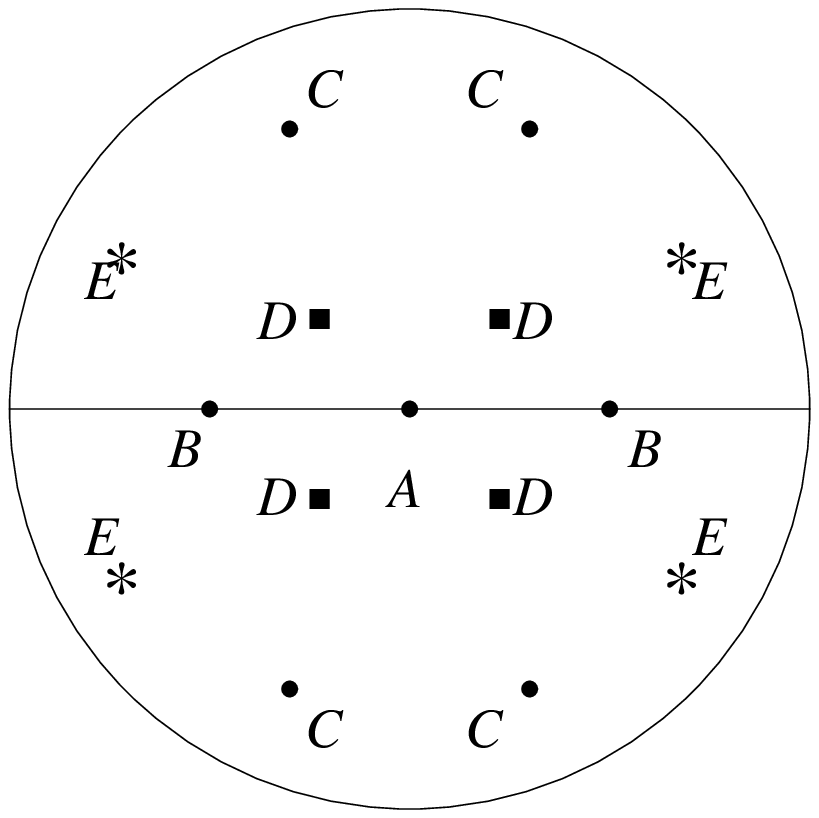,width=6cm}
\hspace{4cm} \epsfig{figure=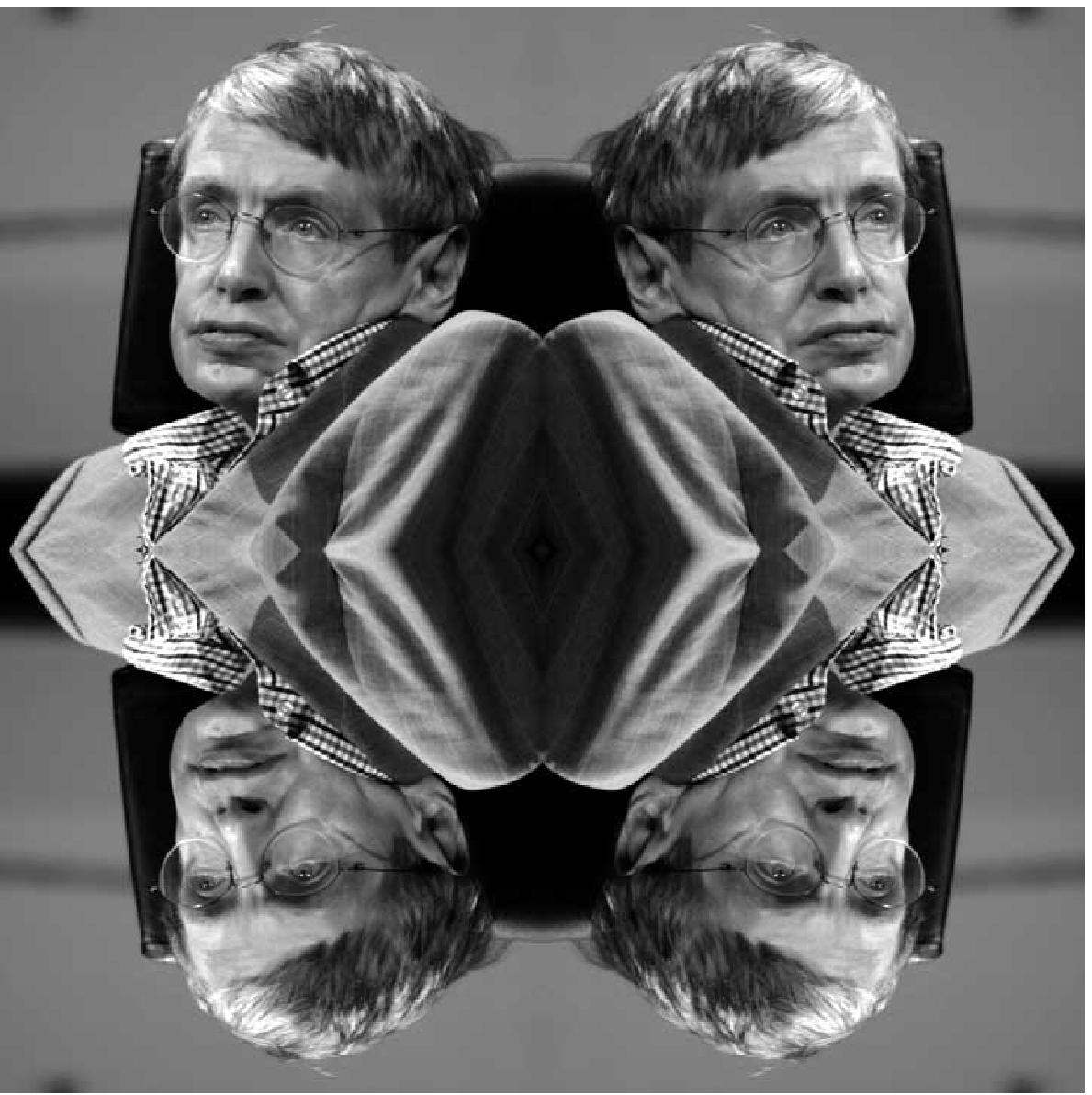,width=6cm}
\end{tabular}
\caption{\label{fig1}\footnotesize There is four absolutely same
copies of a picture on the area. This feature makes decreasing of
the effective area containing the information,
$A_{effective}=\frac{1}{4}A$. The direct consequence is that
supposing each $1\times \ell_P^2$ holds \textit{one} bit of
information predicts the correct entropy-area relation for black
holes. This is indebted to self-relative information paradigm.}
\end{figure}

\subsection{Picking up (counting) all distinguishable states}
To do more on the third choice we will use the above proposal of
removing ambiguities from the choices as mentioned for the second
choice, $B$. Up to now, we have a point $A$, and two equivalent
$B_1$ and $B_2$. Now to introduce the third point, $C$, with
relative distances $d_1$ and $d_2$ with respect to $A$ and $B$,
there is four different choices as it is obvious in figure 5. All
$C$'s have a distance $d_1$ from $A$ but $C_{1B_1}$ and $C_{2B_1}$
have distance $d_2$ from $B_1$ and $C_{1B_2}$ and $C_{2B_2}$ have
distance $d_2$ from $B_2$. So now we have four indistinguishable
states, $\{A,B_1,C_{1B_1}\}$, $\{A,B_1,C_{2B_1}\}$,
$\{A,B_2,C_{1B_2}\}$ and $\{A,B_2,C_{2B_2}\}$. It means that these
four sets have the same information and as a consequence they must
be counted once in our method. Exactly, similar to picking up $B$ to
remove any ambiguities we will take all of these four
indistinguishable states. Now for the fourth choice, $D$, with
(allowable) relative distances $d'_1$, $d'_2$ and $d'_3$ to $A$, $B$
and $C$ respectively one must choose one of the equivalent
sets\footnote{We stress on allowable to make the existence of $D$
possible on the two-dimensional surface, i.e. three circles with
origins $A$, $B$ and $C$ and radii $d'_1$, $d'_2$ and $d'_3$
respectively, must have an intersection to make the radii
allowable.}. But the interesting feature is that now for each set
only one choice exists since there is only \textit{one} intersection
point for three circles generally. So for each set we have a $D$ and
in total we have four $D$'s and similarly for next choices. It means
that to ignore the ambiguities all the points must appear four times
in the area or in other words, the area has four similar copies of
one sequence. It means that the effective area is
$A_{effective}=\frac{1}{4}A_{BH}$, figure 6. And since $Area \propto
N^2$ then the effective information-filled bits are not $N^2$ but
are $\frac{1}{4}N^2$. The entropy due to the relations
$S=\log{2^{\frac{1}{4}N^2}}=(\log{2})\times\frac{1}{4}N^2$ and
$A_{BH}=N^2\ell_P^2$ becomes $S_{BH}=
\frac{1}{4}\frac{A_{BH}}{\ell_P^2}$ with exactly\footnote{We ignore
a $\log 2$ factor since it does not contribute to our discussion. As
mentioned in the footnote \ref{footnote} this factor is a natural
factor due to our binary structure of bits.} the correct factor
$\frac{1}{4}$. Note that deducing of this factor is a direct
consequence of self-relative viewpoint on the information.

\subsection{The logarithmic correction term}
Due to the above discussions, the idea of self-relative information
can show why the entropy-area relation of black holes may be correct
even with assuming $\ell_P^2$ as a holder of \textit{one} bit of
information. But naturally this way of looking at the problem makes
the correction terms appearance spontaneously and this is an
advantage that this method has. As mentioned above, in choosing the
first point $A$, there is not $N^2$ choices but only one choice
because in the absence of any ticked bit there is no differences
between the bits. To reduce this degeneracy the total number of
states must be divided by $N^2$. And similarly for the second point,
$B$, we are allowed to choose only $\sim N$ states instead of $\sim
N^2$. Because for two points only the relative distance is important
in self-relative information method. It means that the total number
of distinguishable states must be divided by $N^3$. So the entropy
becomes $S_{BH}=\log {\frac{2^{\frac{1}{4}N^2}}{N^3}}$ that is
$S_{BH}=\frac{1}{4}N^2-\frac{3}{2}\log{N^2}$ and since
$A_{BH}=N^2\ell_P^2$ then
$S_{BH}=\frac{1}{4}\frac{A_{BH}}{\ell_P^2}-\frac{3}{2}\log{\frac{A_{BH}}{\ell_P^2}}$.
This result is totally in agreement with the previous results. The
constant factor of the logarithmic term is exactly in agreement with
other approaches \cite{string bh,terno}.

Note that the above results are not restricted to binary bit
concept. For example for a $d$-level system, i.e. each bit can have
$d$ independent state, the entropy will be
$S_{BH}=\log\frac{d^{\frac{1}{4}N^2}}{N^3}$ and consequently
$S_{BH}=\frac{1}{4}N^2 \log d-\frac{3}{2}\log{N^2}$ or\footnote{As
mentioned in \cite{shanon} the basis of the logarithmic function in
definition of entropy by Shanon law is an arbitrary and we have
fixed it $d$ itself.}
$S_{BH}=\frac{1}{4}\frac{A_{BH}}{\ell_P^2}-\frac{3}{2}\log_d{\frac{A_{BH}}{\ell_P^2}}$.
It is worth mentioning that the choice of the logarithmic function's
basis does not affect the coefficient of this term. Since as
assigned before this factor comes from avoiding any ambiguities to
pick up the first and the second points and it is a natural factor
due to two-dimensionality of black hole area.

\section{Conclusions}
In this paper, in a model independent information based method we
have deduced the entropy-area relation for a black hole which not
only illustrates the correct coefficient in linear term, i.e.
$\frac{1}{4}$, but also predicts a logarithmic correction term
naturally. To do this, we have introduced a new interpretation of
the concept of accessible information in a sequence of bits. The
idea is based on the reality that in the absence of any reference
dictionary or database, decoding a sequence of bits is impossible.
And as a consequence, there is no pure knowledge about the
information carried by that sequence. So to count the information in
such a sequence the usual method of counting leads to incorrect
results because the existence of the dictionary defines the meaning
of the sequences. To remove this problem, we have proposed that all
the information must be held by each sequence itself. That means the
structure of the sequence itself must show all the information about
that sequence. One possibility is that, the relative place of the
bits in a same sequence are understandable and physical. This
viewpoint on the information makes requiring to a dictionary
unnecessary. In other words only the distinguishable internal
relative structure of sequences leads to independent sequences. This
feature is very important for counting the distinguishable
sequences. We have named the method self-relative information
viewpoint\footnote{This method of thinking is exactly similar to
Mach's thinking about the geometry.} since in this method all the
requisites are the sequences themselves and only the relative
positions of the bits in a given sequence have information.

It is shown in the context of self-relative information viewpoint,
that the very famous relation of entropy-area for a black hole can
be deduced i.e. $S_{BH}=\frac{1}{4}A_{BH}$. In standard viewpoint on
the black hole entropy-area relation there is a $\frac{1}{4}$
proportionality factor that results in each $4 \times \ell_P^2$
element holding \textit{one} bit of information. The natural
question is that why $4 \times \ell_P^2$ and not $1 \times \ell_P^2$
is the fundamental holder of information? We have shown that by
counting the distinguishable possibilities using the self-relative
method, not only is $1 \times \ell_P^2$ the fundamental holder of
the information but also $S_{BH}=\frac{1}{4}A_{BH}$ is
valid\footnote{And even the logarithmic correction term appears
naturally.}. Note that for holding \textit{one} bit of information,
$1 \times \ell_P^2$ is more natural and credible than $4 \times
\ell_P^2$. To do calculations we started with inverse method which
constructed all the distinguishable states and then count them using
the self-relative information paradigm. Another point is that during
the calculations we must try to remove the possible ambiguities to
obtain the correct final answer. The logarithmic correction term
appears naturally as a direct consequence of the method that can be
interpreted as an evidence for the legitimacy of this way of
thinking. It is necessary to say some words about the first two
chosen points. All points are on a two-dimensional surface i.e. the
area of the black hole. To begin the counting we suppose that all
the points are $0$ i.e. a white two-dimensional space. Then we
picked up\footnote{I.e. changing the value from $0$ to $1$ or the
color from white to black.} the first point, $A$, on this
two-dimensional area. But the universe for this first black point in
the white area is zero-dimensional because it cannot understand the
dimensionality of the area with any experiments\footnote{Do not
forget that all the physical quantities are relative and for the
first point there is no other points for doing any physical
comparison even understanding the dimension of space.}. This feature
is essential in calculating the logarithmic correction. Since to
choose a point in zero-dimensional space there is no $N^2$ choices
but one choice even if the space is two-dimensional\footnote{Note
that we have supposed $A_{BH}=N^2 \ell_P^2$.}. This is because of
the blindness of a sole point to the dimensionality of its perimeter
space. Or in other words, for the first choice there is no
difference between all the points or all the points are equivalent.
To continue the counting we picked up the second black point, $B$.
Now there is only two black points, $A$ and $B$. For these two
points only their relative distance is meaningful and physical since
this is the only relative quantity for a space with only two
objects. Two points build a one-dimensional space and they are blind
to any extra dimension, on the black hole's area i.e. the second
dimension. It means that two points see the space, the area of
interest, only in a one-dimensional form and not two-dimensional
even if it is the dimensionality in reality. So the choices for the
second point is not proportional to $N^2$ but it has only $N$
choices due to one-dimensionality for only two points. The procedure
for the next points becomes trivial since introducing the third
point on the area of the black hole makes the space's dimensionality
(the black hole's area) recognizable and therefore the
dimensionality of the area becomes physical for the third point so
the choices are proportional to $\sim N^2$. For the next points the
area is two-dimensional since the third point has established the
dimensionality of the space for all the next points. This is the
reason for dividing the total number of choices by $N^3$ which
causes the logarithmic correction term to appear.

The self-relative information proposal can be seen in the context of
loop quantum gravity approach due to similar structures in some
senses. The entropy-area relation has been considered in the latter
approach as mentioned in \cite{loop bh}. In this scope as considered
in \cite{rovelli} different states represent status of a black hole
which are equivalent if and only if be indistinguishable by
measurements outside the black hole region. That is, the information
on the horizon and not inside it, is considerable
\cite{terno,rovelli}. It is exactly what is done in self-relative
approach. Also, in comparison to \cite{terno}, which is a quantum
informational approach to black hole entropy-area relation, an
interesting point is similar prediction for the coefficient factor
of logarithmic correction term, $-\frac{3}{2}$, notwithstanding
arbitrary level freedom for each bit. In \cite{terno} this universal
factor is a consequence of entangled qubits but in our case it is a
consequence of two-dimensionality of area. Another point is that the
method in \cite{terno} cannot fix the coefficient of linear term if
$1\times\ell_P^2$ would assumed as the fundamental area that is a
vital difference to the self-relative approach.

It is worth mentioning that to reconfirm the self-relative
information paradigm it is possible to check it with the results
from multi-dimensional models. They have shown that the entropy-area
relation for black holes embedded in a $D$-dimensional geometry is
same as the above result for the four-dimensional geometry with a
$\frac{1}{4}$ factor. The application of the self-relative
information proposal to this multi-dimensional configuration is not
very straightforward. The first steps are same as four-dimensional
one i.e. choosing the points is in the same manner as above. But
there is a crucial interpretation that is the area holds information
e.g. $1\times \ell_P^2$ holds \textit{one} bit of information. This
concept is very crucial in the counting method. Unfortunately, this
interpretation is not very obvious when selected points are not
associated with area but with super-area\footnote{By super-area we
mean an object with more than two dimensions that plays the same
role as the ordinary area in the calculation of entropy-area
relation in four-dimensional geometry with a two-dimensional
horizon.}. It seems natural that when a fundamental area holds
\textit{one} bit of information then a fundamental super-area holds
more\footnote{It makes a factor greater than $1$ in the entropy-area
relation that is in agreement with the self-relative information
paradigm. It is very simple to show that this approach predicts
$\frac{1}{2^{D-2}}$ for a black hole embedded in a $D$-dimensional
geometry which is less than $\frac{1}{4}$. This brings some hope for
self-relative information paradigm to be correct even for
$D$-dimensional geometry with an explicit definition for how much
information exists in a fundamental super-area.}. In this sense this
problem is still open to interpretation\footnote{We would like to
point that even if this approach does not work for $D$-dimensional
geometries it still is interesting since our universe has four
macroscopic dimensions. Maybe for calculation in extra dimensions
those are not observables at least macroscopically and hence we need
new definitions and notions for defining information. One suggestion
can be quantum information since usually these extra dimensions
correspond to quantum geometrical effects.}. Also the application of
self-relative information proposal should be used for other
four-dimensional black holes i.e. rotating and charged black holes.
Although the above open problems exist but perhaps considering the
self-relative information paradigm as a way to understand better the
entropy-area relation can help us find the rest of the iceberg of
quantum gravity \cite{strominger}.

Finally, we would like to mention that among different approaches to
quantization of general relativity like string theory or loop
quantum gravity etc. there are some common features. As mentioned in
\cite{smolin}, discreetness of geometrical objects (such as length,
area and volume) and the holographic principle are common in
different approaches to quantum gravity. The idea introduced in
\cite{smolin} says that to study the true quantum gravity, one must
assume these features as the initial axioms and build the theory on
these bases. Then in the semi-classical limits recover classical
general relativity or quantum mechanics. We would like to suggest
that self-relative thinking about the information can be another
essential axiom about the nature. This idea is amplified by
mentioning that there is no dictionary to decode the nature's
information so the only way to think about it, is self-relative
viewpoint. For the final words it seems good to note that if
somebody believes in ``it from bit" idea of Wheeler \cite{wheeler}
then self-relative information plays a crucial role to interpret the
quotation.

\section{Summary}
In this short section we will briefly present our paradigm in an
axiomatic way.\\
\\
AXIOM I: Discreetness of the area.\\
Evidence I: Existence of the fundamental area, e.g. $\ell_P^2$, is a
common sense in quantum theories of gravity\footnote{There is a
question that is $\ell_P$ the fundamental length? or for example
$k\times \ell_P$ is the fundamental one? where $k$ is a
proportionality factor. The answer is not straightforward. But in
the absence of any knowledge on this problem proposition of $\ell_P$
as the fundamental length is natural. In other viewpoint,
combination of the present self-relative information proposal and
the entropy-area relation of black holes can be interpreted as
an evidence for being $\ell_P$ as the fundamental length.}.\\
\\
AXIOM II: Each $1\times \ell_P^2$ holds \textit{one} bit of
information.\\
Evidence II: Maybe ``it from bit"\footnote{For our proposal this
proposition is only a way to interpret the
concept of the entropy especially for black holes.}.\\
\\
AXIOM III: Self-relative (Machian) information paradigm.\\
Evidence III: There is no external knowledge about the
information (i.e. there is no dictionary or database).\\
Evidence III$'$: The notion of relativity is an essential concept in
definition of physical quantities.\\
\\
Black hole entropy-area THEOREM: From the AXIOMS I, II and
III\footnote{It is worth mentioning that AXIOM III (self-relative
information) is very crucial to prove this theorem.} it can be shown
that the following relation\footnote{Note that the existence of
logarithmic correction term is a natural consequence of this
method.} in Planck's units exists between entropy of a black hole,
$S_{BH}$, and its horizon area, $A_{BH}$,
\begin{eqnarray}\nonumber
S_{BH}=\frac{1}{4}A_{BH}-\frac{3}{2}\log A_{BH}.\nonumber
\end{eqnarray}

\vspace{10mm}\noindent\\
{\bf Acknowledgments}\vspace{2mm}\noindent\\ I am grateful to M.
Asoudeh, N. Doroud, G. W. Gibbons, H. Gourani, G. R. Jafari, S.
Jalalzadeh, N. Rafiei, H. R. Sepangi, M. M. Sheikh-Jabbari and J. P.
van der Schaar for fruitful discussions and their comments. I would
like to thank H. R. Sepangi also for careful reading of the
manuscript.

\end{document}